\documentclass[twocolumn,showpacs,amsmath,amssymb]{revtex4-1}
\usepackage{graphicx}
\usepackage{color}
\usepackage[colorlinks,allcolors=blue]{hyperref}

\begin{document}

\title{Chern insulator in a ferromagnetic two-dimensional electron system with Dresselhaus spin-orbit coupling}

\author{Rui-An Chang$^1$}

\author{Ching-Ray Chang$^{2,3}$}

\affiliation{$^1$Graduate Institute of Applied Physics, National Taiwan University, Taipei 10617, Taiwan\\
	$^2$Department of Physics, National Taiwan University, Taipei 10617, Taiwan\\
$^3$Center for Theoretical Physics, National Taiwan University, Taipei 10617, Taiwan}

\begin{abstract}
We propose a Chern insulator in a two-dimensional electron system with Dresselhaus spin-orbit coupling, ferromagnetism, and spin-dependent effective mass. The analytically-obtained topological phase diagrams show the topological phase transitions induced by tuning the magnetization orientation with the Chern number varying between $1,0,-1$. The magnetization orientation tuning shown here is a more practical way of triggering the topological phase transitions than manipulating the exchange coupling that is no longer tunable after the fabrication of the system. The analytic results are confirmed by the band structure and transport calculations, showing the feasibility of this theoretical proposal. With the advanced and mature semiconductor engineering today, this Chern insulator is very possible to be experimentally realized and also promising to topological spintronics. 
\end{abstract}

\maketitle

\section{Introduction}\label{introduction}
The quantum Hall effect (QHE) can be observed in two-dimensional electron systems with a strong perpendicular magnetic field which breaks the time-reversal symmetry (TRS). Reseachers also proposed the QHE without a magnetic field in which the TRS breaking is alternatively achieved by magnetic materials. It is the so-called quantum anomalous Hall (QAH) effect, also known as Chern insulators. QAH systems have been intensively studied for their fruitful physics and promising applications in future technology \cite{Haldane,RevModPhys.82.3045,QAHtheory,QAHexp,Wang,QWZ,PhysRevLett.111.086803,PhysRevLett.113.016801,PhysRevLett.113.136403,PhysRevLett.111.136801,PhysRevB.82.161414,PhysRevB.85.045445}. They attract the attention of researchers because of the topologically nontrivial chiral edge states in the absence of external magnetic field, which is important to the development of next-generation electronic devices. QAH states in proximity to superconductors, the chiral topological superconductors \cite{RevModPhys.83.1057,PhysRevB.82.184516,PhysRevB.97.104504,PhysRevB.98.165439,PhysRevB.97.081102,PhysRevB.96.224514,PhysRevB.95.245433}, also attract great attention because of the realization of Majorana fermions \cite{Majorana2008}. The first model of a QAH system is constructed by F. D. M. Haldane in the honeycomb lattice in 1988 \cite{Haldane}. A magnetic topological insulator (TI) was theoretically predicted to be a possible QAH system \cite{QAHtheory} and was later experimentally confirmed in a magnetic TI thin film \cite{QAHexp,Wang}.     

\begin{figure} 
	\includegraphics[scale=0.45]{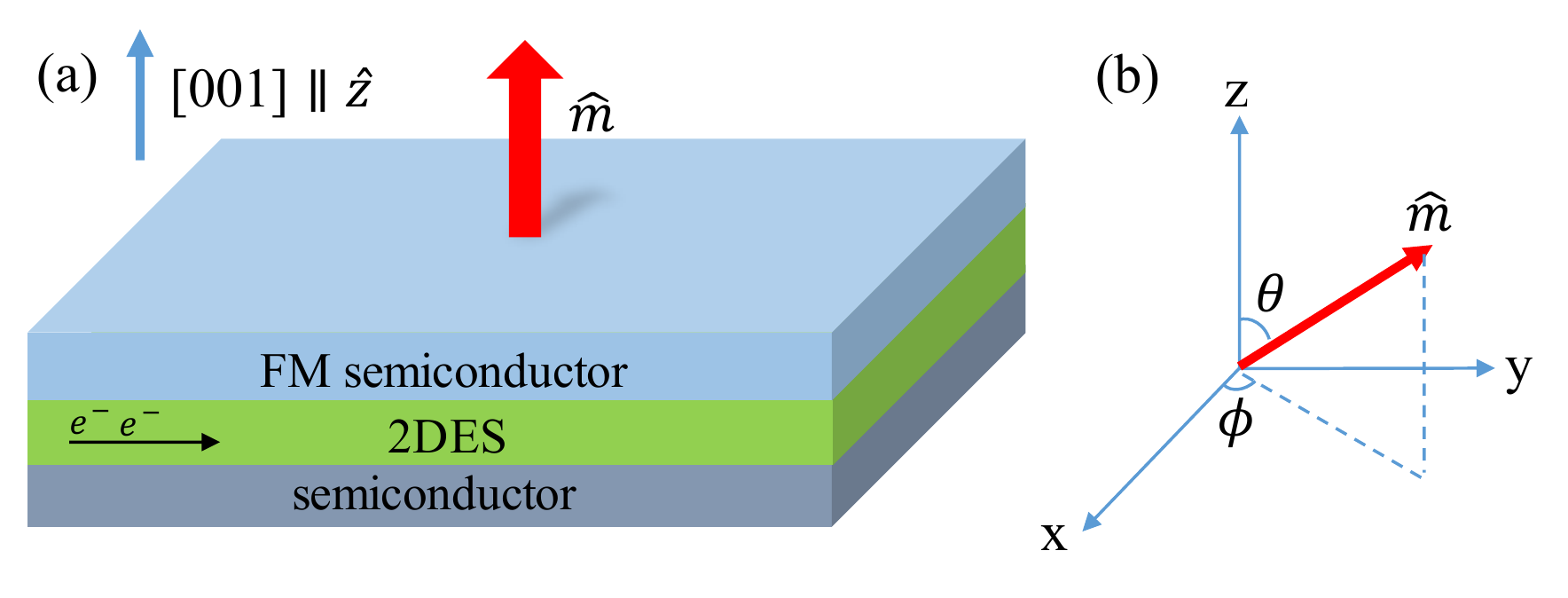} 
	\caption{\label{setup}(a) The 2DES is embedded in the interface of the FM semiconductor/nonmagnetic semiconductor heterostructure. Actually, the 2DES only has a narrow size in the $z$ direction, but it has been specially enlarged for clearness. 
	(b) $\theta$ is the polar angle and $\phi$ is the azimuthal angle. Generically $\hat{\boldsymbol{m}}=(\sin\theta\cos\phi,\sin\theta\sin\phi,\cos\theta)$.}
\end{figure}

In this paper, we propose to realize a Chern insulator in a two-dimensional electron system (2DES) embedded in the interface of a semiconductor heterostructure which is manufactured by growing a zinc-blende ferromagnetic (FM) semiconductor, such as (In,Fe)As \cite{FMS_Nat,FMS_Appl}, on another zinc-blende nonmagnetic semiconductor, as shown in Fig.~\ref{setup}(a). Because the Dresselhaus spin-orbit coupling (SOC) \cite{Winkler} is proved to be present in semiconductors with the zinc-blende crystal structure \cite{Dresselhaus}, a 2DES formed by such a semiconductor heterostructure also has the Dresselhaus SOC. Besides, by properly orienting the lower nonmagnetic semiconductor and controlling the growing direction of the upper magnetic one such that their individual [001] directions are all aligned to the $z$ direction shown in Fig.~\ref{setup}(a), the 2DES system embedded in the interface has a normal vector in the [001] direction. Thus, the Dresselhaus [001] SOC is present in the 2DES.
Furthermore, by the proximity effect of the FM semiconductor, the 2DES has an exchange coupling to it. So the required TRS breaking in this QAH system is achieved by the FM ordering in the FM semiconductor. 

In addition to the Dresselhaus [001] SOC and the exchange coupling to the FM semiconductor, there is still another physical phenomenon required to be considered in this 2DES, the so-called spin-dependent effective mass. It could be easily and literally understood that spin-up and spin-down electrons have different effective masses. This effect has been experimentally verified in 2DESs \cite{2DEG,magnetic_2DEG}. In the 2DES of this paper, the spin-polarization in the FM semiconductor permeating into the 2DES through the proximity effect makes spin-up and spin-down electrons have different effective masses \cite{spin-dependent} such that the effect of spin-dependent effective mass, intrinsically present in a nonmagnetic 2DES \cite{2DEG}, can be more salient in our ferromagnetic 2DES. Because mass is nothing but inertia, different effective masses will make spin-up and spin-down electrons have different kinetic hoppings.

Combining the Dresselhaus [001] SOC, exchange coupling, and spin-dependent effective mass, it will be shown that such a 2DES is a Chern insulator which shows the QAH effect. Meanwhile, inspired by recent spintronic research \cite{Blugel,Nature_STT}, the topological phase transitions will be induced by tuning the magnetization orientation [Fig.~\ref{setup}(b)], which is more practical than traditionally manipulating the exchange coupling strength which is no longer a tunable parameter after the fabrication of the system.

\section{model and topological invariant}\label{II}
From Sec. \ref{introduction}, the Hamiltonian should include the terms derived from the Dresselhaus [001] SOC, exchange coupling to the FM semiconductor, and the kinetic hoppings with the effect of spin-dependent effective mass. 
The Dresselhaus [001] Hamiltonian is $(\beta/\hbar)(p_{x}\sigma_{x}-p_{y}\sigma_{y})$ \cite{Dresselhaus,Dresselhaus[001]} in which $\beta$ is the Dresselhaus coupling constant. Now we do the substitution $p_{x}\rightarrow -i\hbar\;\partial/\partial_{x}$ and $p_{y}\rightarrow -i\hbar\;\partial/\partial_{y}$ in $H_{D}\psi(x,y)$ and use the finite difference method to discretize the 2DES into a square lattice with the lattice constant $a$. Under this case, 
\begin{eqnarray}
p_{x}\psi(x,y)\approx (-i\hbar/2a)[\psi(x+a,y)-\psi(x-a,y)]\nonumber\\
p_{y}\psi(x,y)\approx (-i\hbar/2a)[\psi(x,y+a)-\psi(x,y-a)]
\end{eqnarray}
so the real-space tight-binding Dresselhaus [001] Hamiltonian with the nearest-neighbor hopping can be easily read out as follows
\begin{eqnarray}\label{real-space_Dresselhaus}
H_{D}=\sum_{i}&&c^{\dagger}_{i+x}(-\frac{i}{2}t_{D}\sigma_{x})c_{i}+c^{\dagger}_{i+y}(\frac{i}{2}t_{D}\sigma_{y})c_{i}+h.c.\;,
\end{eqnarray}
where $c_{i}=(c_{i\uparrow}\;c_{i\downarrow})^{\rm T}$ and $t_{D}=\beta/a$. $c^{\dagger}_{i\sigma}$ and $c_{i\sigma}$ are creation and annihilation operators of an electron with spin $\sigma=\uparrow,\downarrow$ on the $i$\textsuperscript{th} site of the square lattice. $i+x$ and $i+y$ denote the nearest-neighbor sites in the $+x$ and $+y$ directions relative to the $i$\textsuperscript{th} site, respectively.

In the 2DES discretized as a square lattice, the kinetic Hamiltonian originally is
\begin{eqnarray}\label{real-space_kinetic}
H_{K}=\sum_{i}c^{\dagger}_{i+x}(\frac{t}{2}\textbf{I})c_{i}+c^{\dagger}_{i+y}(\frac{t}{2}\textbf{I})c_{i}+h.c.\;,
\end{eqnarray}
where $t$ is the kinetic hopping and $\textbf{I}$ is a two-by-two identity matrix whose diagonal elements represent the kinetic hoppings of spin-up and spin-down electrons, respectively. However, as mentioned in the Sec.~\ref{introduction}, spin-up and spin-down electrons have different kinetic hoppings, so the hopping term $t\textbf{I}$ should be modified as follows:
\begin{eqnarray}
t\textbf{I}\rightarrow
\begin{pmatrix} 
t+t' & 0 \\
0 & t-t' 
\end{pmatrix}=t\textbf{I}+t'\sigma_{z},
\end{eqnarray}
in which $t'$ represents the hopping energy splitting between spin-up and spin-down electrons. Thus, the kinetic Hamiltonian should be
\begin{eqnarray}\label{real-space_kinetic_2}
H_{K}=&&\frac{1}{2}\sum_{i}c^{\dagger}_{i+x}(t\textbf{I}+t'\sigma_{z})c_{i}+c^{\dagger}_{i+y}(t\textbf{I}+t'\sigma_{z})c_{i}\nonumber\\&&+h.c.
\end{eqnarray}
Along with the Zeeman term derived from the exchange coupling to the FM semiconductor
\begin{eqnarray}\label{Zeeman}
H_{Z}=c^{\dagger}_{i}(\Delta\hat{\boldsymbol{m}}\cdot\boldsymbol{\sigma})c_{i}
\end{eqnarray}
where $\Delta$ is the exchange coupling strength, $\hat{\boldsymbol{m}}=(\sin\theta\cos\phi,\sin\theta\sin\phi,\cos\theta)$, and $\boldsymbol{\sigma}=(\sigma_{x},\sigma_{y},\sigma_{z})$, the total real-space tight-binding Hamiltonian is
\begin{eqnarray}\label{real-space}
H=H_{D}+H_{K}+H_{Z}.
\end{eqnarray}
The band structure calculation can be performed by doing the Fourier transform to this Hamiltonian along the direction with the periodic boundary condition. The results will be shown in the Sec.~\ref{TPT}.

So far, we have successfully constructed the Hamiltonian of this 2DES. In order to discuss the topological properties of this system, we do the Fourier transform of Eq.~(\ref{real-space}) into the $\mathbf{k}$-space
\begin{eqnarray}\label{QWZ2}
H=\sum_{\mathbf{k}}\sum_{\alpha,\beta=\uparrow\downarrow} c^{\dagger}_{\mathbf{k} \alpha} H_{\alpha \beta}(\mathbf{k}) c_{\mathbf{k} \beta},
\end{eqnarray}
where
\begin{eqnarray}\label{QWZ}
H(\mathbf{k})=&&-t_{D}\sin k_{x}\sigma_{x}+t_{D}\sin k_{y}\sigma_{y}\nonumber\\ 
&&+t(\cos k_{x}+\cos k_{y})\textbf{I}+t'(\cos k_{x}+\cos k_{y})\sigma_{z}\nonumber\\&&+\Delta\hat{\boldsymbol{m}}\cdot\boldsymbol{\sigma}.
\end{eqnarray}
$c^{\dagger}_{\mathbf{k} \alpha}$ and $c_{\mathbf{k} \alpha}$ are creation and annihilation operators of an electron with momentum $\mathbf{k}$ and spin $\alpha$.   
          
With $H(\mathbf{k})$, we can investigate the topological property of this system by calculating the topological invariant-Chern number \cite{Chang}
\begin{equation}\label{Chern}
C=\frac{1}{4\pi}\int_{BZ} dk_{x} dk_{y}\;\mathbf{\hat{d}}\cdot(\frac{\partial \mathbf{\hat{d}}}{\partial k_{x}} \times \frac{\partial \mathbf{\hat{d}}}{\partial k_{y}}),
\end{equation}
where $\mathbf{\hat{d}}$ is the unit vector of $\mathbf{d(k)}$ by which $H(\mathbf{k})$ can be written as $H(\mathbf{k})=\mathbf{d(k)}\cdot\boldsymbol{\sigma}$. This integration is defined in the first Brillouin zone (BZ) with $k_{x}$ and $k_{y}$ ranging from -$\pi$ to $\pi$. In addition, if the system is half filled, the quantized Hall conductance can be expressed through the Chern number \cite{TKNN,Chang}
\begin{eqnarray}\label{cond}
\sigma_{xy}=\frac{e^{2}}{h} C\;,
\end{eqnarray} 
where $e$ is the charge of an electron and $h$ is the Planck constant. It is straightforward to find that only the coefficients of $\sigma_{x}$, $\sigma_{y}$, and $\sigma_{z}$ in Eq.~(\ref{QWZ}) can affect the topological property. That is to say, we can throw $t(\cos k_{x}+\cos k_{y})\textbf{I}$ away without changing the topology. Besides, $t'$ is renamed as $t$ for convenience in notation. So the $H(\mathbf{k})$ becomes
\begin{eqnarray}\label{QWZ_new}
H(\mathbf{k})=&&-t_{D}\sin k_{x}\sigma_{x}+t_{D}\sin k_{y}\sigma_{y}+t(\cos k_{x}+\cos k_{y})\sigma_{z}\nonumber\\&&+\Delta\hat{\boldsymbol{m}}\cdot\boldsymbol{\sigma}.
\end{eqnarray}
\begin{figure}
	\centering 
	\includegraphics[scale=0.365]{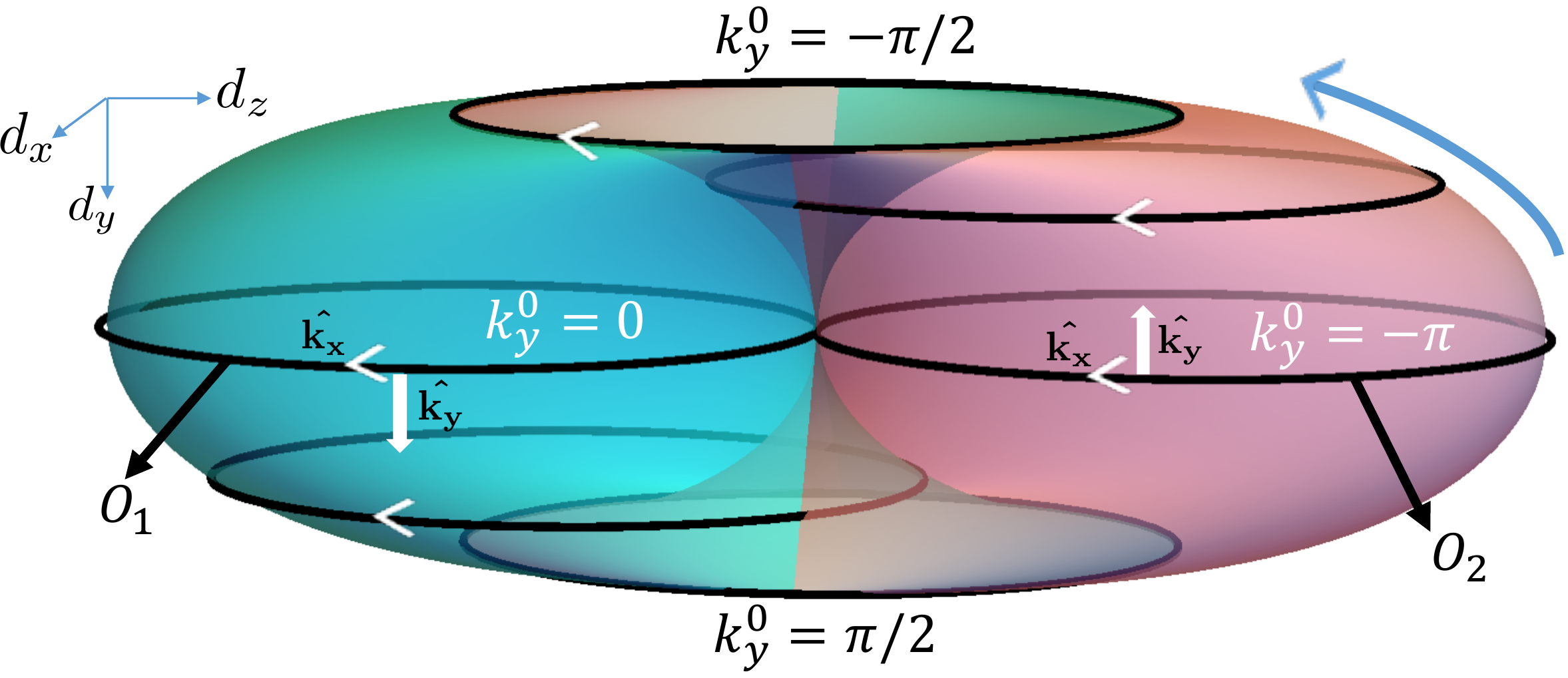}  
	\caption
	{\label{S2}Sphere $S^{2}$ formed by the sweep of $\mathbf{d(k)}$. Starting from the ellipse $O_{2}$ at $k_{y}^{0}=-\pi$, many ellipses of Eq.~(\ref{ellipse1}) can be sequentially traced out along the direction of the curved arrowhead. Viewed from the $-d_{y}$ direction, the parameter $k_{x}$ is evolving clockwise on each ellipse at each given value of $k_{y}=k_{y}^{0}$. Besides, $k_{y}$ is evolving to the $+d_{y}$ direction when it is within $[-\pi/2,\pi/2]$; otherwise, $k_{y}$ is evolving to the $-d_{y}$ direction. With the identification of the $\hat{\mathbf{k_{x}}}$ and $\hat{\mathbf{k_{y}}}$ directions on $S^{2}$, the normal vector is defined by $\hat{\mathbf{k_{x}}}\times\hat{\mathbf{k_{y}}}$ (pointing to the outer side of $S^{2}$).} 
\end{figure}
The same procedure is also applied to $H_{K}$. Back to the discussion of Chern number, the expression in Eq.~(\ref{Chern}) has a direct geometrical interpretation and $\mathbf{\hat{d}(k)}$ totally determines the topological property of this system. The function $\mathbf{\hat{d}(k)}$ represents a mapping from the momentum space (BZ) to the sphere $S^{2}$ in $\mathbf{d}$-space, as shown in Fig.~\ref{S2}. This mapping can be denoted as $\mathbf{\hat{d}(k)}$\;:\;$T^{2}\rightarrow S^{2}$, where BZ $\in$ $T^{2}$ (torus) in topology \cite{QAHreview}.
$(\frac{\partial \mathbf{\hat{d}}}{\partial k_{x}} \times \frac{\partial \mathbf{\hat{d}}}{\partial k_{y}})$ in the integrand of Eq.~(\ref{Chern}) is nothing but the Jacobian of the mapping from BZ to $\mathbf{d}$-space. $\mathbf{\hat{d}}\cdot(\frac{\partial \mathbf{\hat{d}}}{\partial k_{x}} \times \frac{\partial \mathbf{\hat{d}}}{\partial k_{y}})$ along with $dk_{x}dk_{y}$ is therefore an infinitesimal solid angle in $\mathbf{d}$-space mapped from an infinitesimal area $dk_{x}dk_{y}$ in the momentum space (BZ). From this geometrical interpretation, we can know that the integration divided by $4\pi$ is exactly how many times the vector $\mathbf{\hat{d}(k)}$ can wind around the origin as $k_{x}$ and $k_{y}$ run through the whole BZ. Therefore, the Chern number, an integration in two-dimensional BZ, is just the winding number of the vector $\mathbf{\hat{d}(k)}$ in three-dimensional $\mathbf{d}$-space. The geometrical analysis of the Chern number expression is not superfluous but important for appreciating the topological property of this system. As we will see, the topological phase transitions can be identified by the evolution of the sphere $S^{2}$ with respect to a certain degree of freedom such as $\theta$. 

In the next section, we will discuss how the vector $\mathbf{d(k)}$ behaves as $k_{x}$ and $k_{y}$ run through the whole BZ. The topological property of this system depends on if the $\mathbf{d}$-space origin is enclosed in the sphere $S^{2}$ traced out by the sweep of the vector $\mathbf{d(k)}$. This system is topologically nontrivial or trivial if the origin is or is not enclosed. Consequently, a very significant part of our work is to identify the critical state between the trivial and nontrivial phases, where the $\mathbf{d}$-space origin is on the boundary of the sphere $S^{2}$.
\begin{figure}[t]
	\centering 
	\includegraphics[scale=0.45]{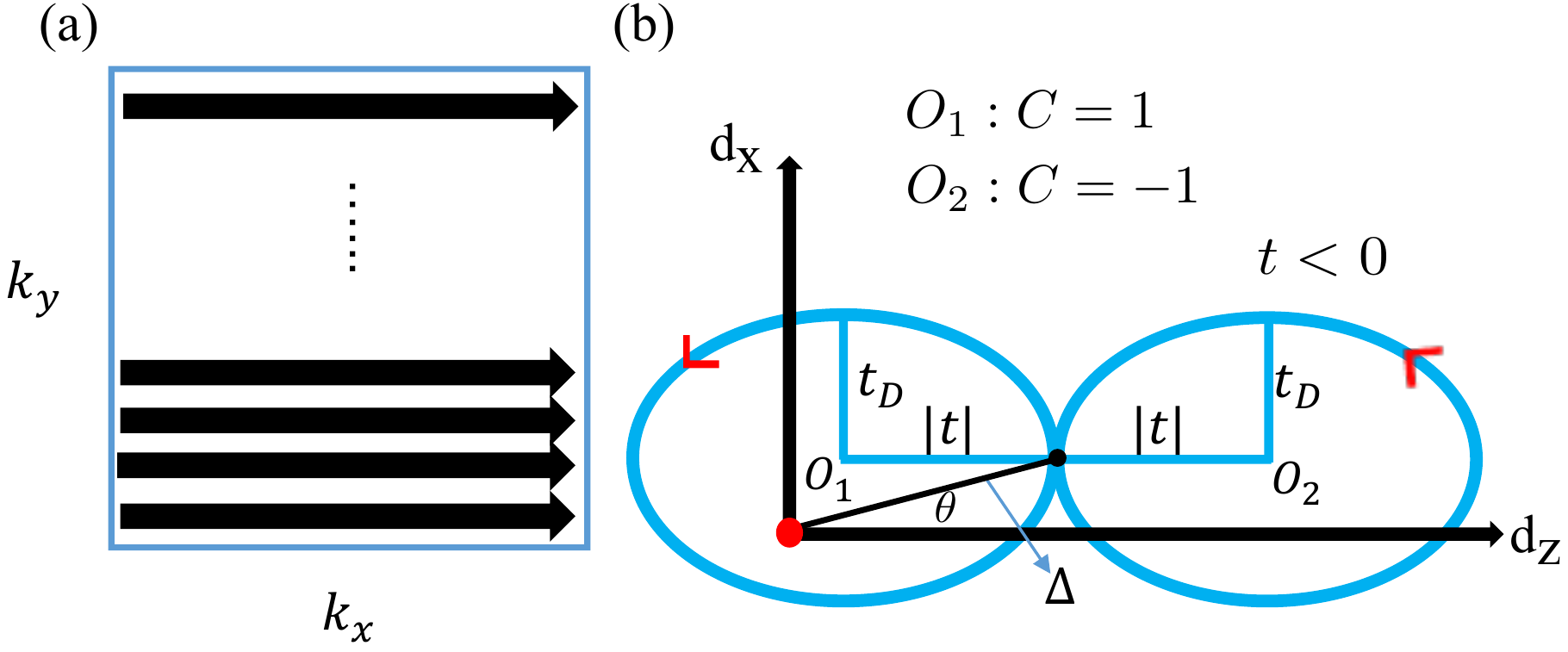}  
	\caption
	{\label{BZ}(a)
		It shows how every point in the BZ is gone through. For a specific value of $k_{y}=k_{y}^{0}$, the line segment $k_{x}\in [-\pi,\pi]$ in the BZ maps to a closed elliptical trajectory in $\mathbf{d}$ space as Eq.~(\ref{ellipse1}). (b) These two tangent ellipses correspond to $k_{y}^{0}=-\pi, 0$. Actually, they are just $O_{2}$ and $O_{1}$ in Fig.~\ref{S2} viewed from the $+d_{y}$-direction. The topological property of this system is determined by the position of the origin (red dot) relative to these two ellipses: $C=1$ or $-1$ if the origin is in $O_{1}$ or $O_{2}$.\;(Note: (b) is not drawn according to the ratios of parameters.) 
	} 
\end{figure} 
\section{topological phase transitions}\label{TPT}\;For the simplification of the following calculation, we assume $\phi$ to be zero such that $\hat{\boldsymbol{m}}$ lies on the $x-z$ plane. So the components of $\mathbf{d(k)}$ extracted from Eq.~(\ref{QWZ_new}) are
\begin{eqnarray}
d_{x}&=&-t_{D}\sin k_{x}+\Delta\sin\theta,\nonumber\\
d_{y}&=&t_{D}\sin k_{y},\nonumber\\
d_{z}&=&t(\cos k_{x}+\cos k_{y})+\Delta\cos\theta.
\end{eqnarray} 
Now we are discussing how the sphere $S^{2}$ would be formed by the sweep of $\mathbf{d(k)}$ as $k_{x}$ and $k_{y}$ run through the whole BZ [Fig.~\ref{BZ}(a)]. For a specific value of $k_{y}=k_{y}^{0}\in[-\pi,\pi]$, an elliptical trajectory in $\mathbf{d}$-space is traced out on the plane of $d_{y}=t_{D}\sin k_{y}^{0}$ with $k_{x}$ varying as a parameter from $-\pi$ to $\pi$. The equation of an ellipse corresponding to a value of $k_{y}=k_{y}^{0}$ is
\begin{equation}\label{ellipse1}
\frac{(d_{z}-\Delta\cos\theta-t\cos k_{y}^{0})^{2}}{t^{2}}+\frac{(d_{x}-\Delta\sin\theta)^{2}}{t_{D}^{2}}=1
\end{equation}  

Other ellipses can be formed under other specific values of $k_{y}^{0}$, as shown in Fig.~\ref{S2}. After every point in the BZ is gone through, a collection of ellipses would be formed with their centers located on the ellipse 
\begin{eqnarray}\label{ellipse2}
	\begin{cases}
		d_{y}=t_{D}\sin k_{y}^{0} \\
		d_{z}=\Delta\cos\theta+t\cos k_{y}^{0}
	\end{cases}
\end{eqnarray}\begin{figure*}
	\centering
	\includegraphics[scale=0.50]{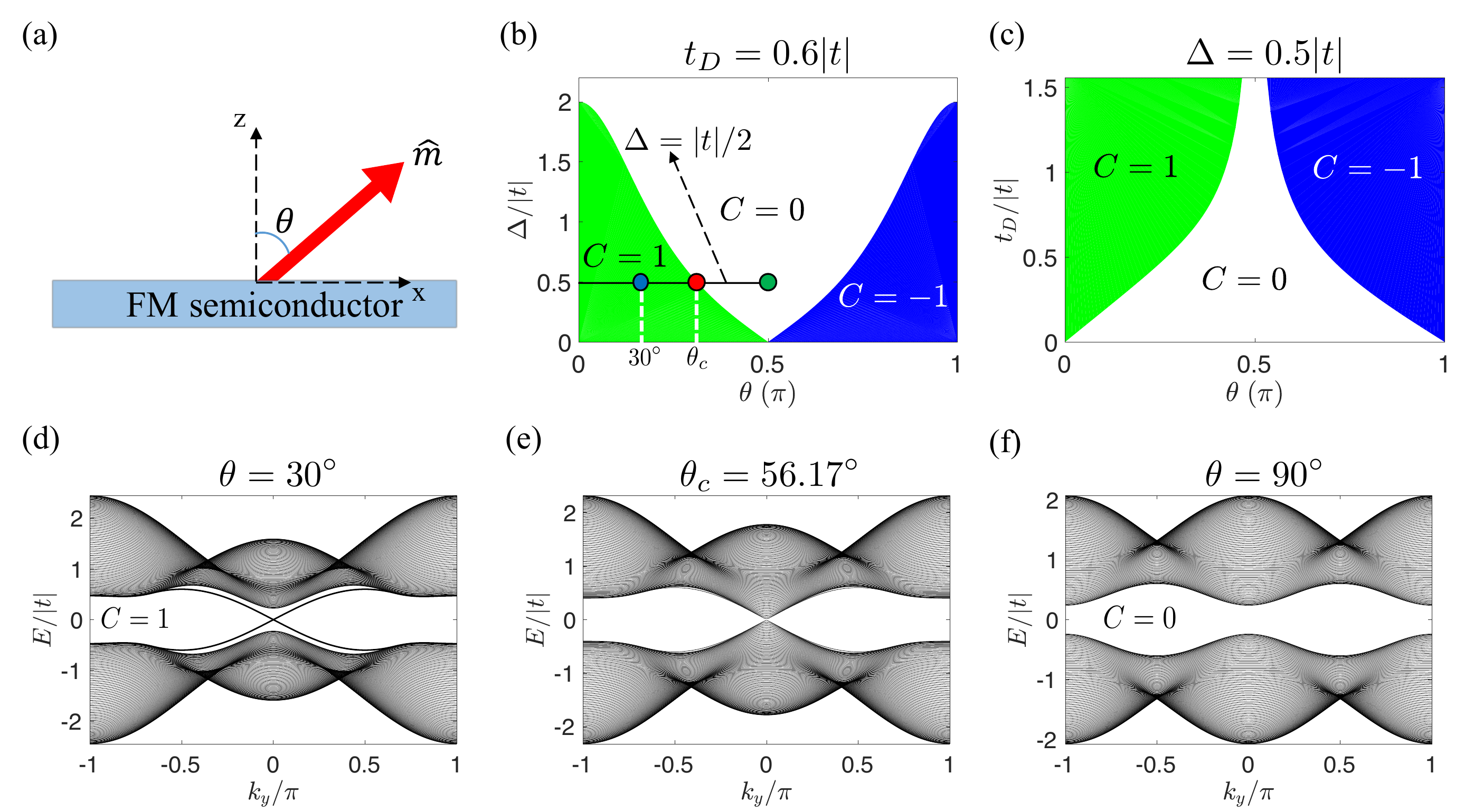}
	\caption{\label{phase}Phase diagrams and energy bands. Topological phase transitions can be induced by the tuning of magnetization orientation $\hat{\boldsymbol{m}}=(\sin\theta,0,\cos\theta)$ in (a). In this figure, $t$ is set to be $-0.5$ eV and other quantities are represented as units of $|t|$. (b) and (c) are phase diagrams of $\Delta-\theta$ and $t_{D}-\theta$ where the alternations of Chern number show the topological phase transitions in terms of $\theta$ (the tuning of $\hat{\boldsymbol{m}}$). In (b) and (c), the green and blue regions represent topologically nontrivial phases where $C=1$ and $C=-1$, respectively. The white regions ($C=0$) stand for the topologically trivial phases. (d), (e), and (f) are energy bands of a strip which has an open boundary condition in the $x$-direction but a periodic boundary condition in the $y$-direction and respectively correspond to three representative values of $\theta$ in (b) under the case that $\Delta=|t|/2$ and $t_{D}=0.6|t|$ (blue, red, and green dots on the black solid line in (b)). In the case of $\theta=30^{\circ}$ (blue dot), chiral edge states are shown in (d) because of the topologically nontrivial phase with $C=1$. In the case of $\theta=\theta_{c}=56.17^{\circ}$ (red dot), (e) corresponds to a topological phase transition because the conduction band and valence band contact with each other. (f) is nothing but a topologically trivial phase (normal insulator) with $\theta=90^{\circ}$ (green dot).}
\end{figure*}, which is on the plane of $d_{x}=\Delta\sin\theta$. In summary, $S^{2}$ is a collection of ellipses of Eq.~(\ref{ellipse1}) whose centers also form an ellipse. By Eq.~(\ref{ellipse2}), it is readily to see that the vector $\mathbf{d(k)}$ can only wrap the origin one time as we go through the whole BZ. Therefore, the Chern number $C$ in Eq.~(\ref{Chern}) can only be 1, 0, or $-1$. 

Because of $\phi$ being set to zero, $S^{2}$ can only be shifted normally to $d_{y}$-direction when the magnetization is being tuned. That is so say, one just has to focus on the intersection of $S^{2}$ with the $d_{y}=0$ plane and observe if the origin is enclosed in $S^{2}$. The intersection is two ellipses corresponding to $k_{y}^{0}=-\pi, 0$ ($d_{y}=0$) in Eq.~(\ref{ellipse1}):
\begin{equation}
	\frac{(d_{z}-\Delta\cos\theta \pm t)^{2}}{t^{2}}+\frac{(d_{x}-\Delta\sin\theta)^{2}}{t_{D}^{2}}=1.
\end{equation}
Actually, they are just $O_{2}$ and $O_{1}$ in Fig.~\ref{S2}, 
whose centers are $(\Delta\cos\theta-t,\Delta\sin\theta)$ and $(\Delta\cos\theta+t,\Delta\sin\theta)$ on the $d_{z}$-$d_{x}$ plane, respectively, as shown in Fig.~\ref{BZ}(b). If the origin is in $O_{1}$ or $O_{2}$, the Chern number $C$ is equal to $1$ or $-1$ because an infinite line from the origin in $O_{1}$ to the negative $z$ direction or that from the origin in $O_{2}$ to the positive $z$ direction will inevitably pierce the inner or the outer surface of $S^{2}$, respectively; otherwise, $C$ is just zero when the origin is not enclosed \cite{shortcourse} (the definition of surface orientation is presented in the caption of Fig.~\ref{S2}). 

In the following, the mathematical requirements of topological phase transitions will be derived in order to obtain the phase diagrams.
For the case of the origin in $O_{1}$ of Fig.~\ref{BZ}(b), we have
\begin{equation}\label{insideO1}
	\frac{(\Delta\cos\theta-|t|)^{2}}{t^{2}}+\frac{(\Delta\sin\theta)^{2}}{t_{D}^{2}}\leq1,\;\;\;\;0<\theta<\frac{\pi}{2}.
\end{equation} 
For the case of the origin in $O_{2}$ of Fig.~\ref{BZ}(b), we have
\begin{equation}\label{insideO2}
	\frac{(\Delta\cos\theta+|t|)^{2}}{t^{2}}+\frac{(\Delta\sin\theta)^{2}}{t_{D}^{2}}\leq1,\;\;\;\;\frac{\pi}{2}<\theta<\pi.
\end{equation}
We expect to obtain the relation between $t_{D}$ and $\theta$. After doing some algebra we can get
\begin{equation}\label{tDtheta1}
	t_{D}\geq\frac{\Delta |t|\sin\theta}{\sqrt{\Delta \cos\theta(2|t|-\Delta\cos\theta)}},\;\;\;\;0<\theta<\frac{\pi}{2}
\end{equation}
for the origin in $O_{1}$ ($C=1$) and
\begin{equation}\label{tDtheta2}
	t_{D}\geq\frac{\Delta |t|\sin\theta}{\sqrt{-\Delta \cos\theta(2|t|+\Delta\cos\theta)}},\;\;\;\;\frac{\pi}{2}<\theta<\pi
\end{equation}
for the origin in $O_{2}$ ($C=-1$). Please note that $2|t|$ needs to be larger than $\Delta |\cos\theta|$ inside the square roots of Eq.~(\ref{tDtheta1}) and~(\ref{tDtheta2}). Indeed, the requirement that $2|t|>\Delta |\cos\theta|$ is necessary to be satisfied or the origin is impossible to be enclosed in $O_{1}$ or $O_{2}$ in Fig.~\ref{BZ}(b). From Eq.~(\ref{tDtheta1}) and~(\ref{tDtheta2}), we can respectively obtain the green ($C=1$) and blue ($C=-1$) regions of the phase diagram that shows the relation between $t_{D}$ and $\theta$, as shown in Fig.~\ref{phase}(c). Topological phase transitions can be induced by the tuning of magnetization orientation at a certain value of $t_{D}$. In this case, the origin can be made to move into or out of the sphere $S^{2}$ shown in Fig.~\ref{S2}. This is the topological mechanism to explain why the topological phase transition can be induced by tuning the magnetization orientation. Following the same procedure, the relation between $\Delta$ and $\theta$ can also be obtained starting from Eq.~(\ref{insideO1}) and~(\ref{insideO2}):
\begin{equation}\label{Deltheta1}
	\Delta\leq\frac{2t^{2}_{D}|t|\cos\theta}{t^{2}_{D}\cos^{2}\theta+t^{2}\sin^{2}\theta},\;\;\;\;0<\theta<\frac{\pi}{2}
\end{equation}
for the origin in $O_{1}$ ($C=1$) and
\begin{equation}\label{Deltheta2}
	\Delta\leq\frac{-2t^{2}_{D}|t|\cos{\theta}}{t^{2}_{D}\cos^{2}\theta+t^{2}\sin^{2}\theta},\;\;\;\;\frac{\pi}{2}<\theta<\pi
\end{equation}
for the origin in $O_{2}$ ($C=-1$). The phase diagram Fig.~\ref{phase}(b) can also be obtained according to Eq.~(\ref{Deltheta1}) and~(\ref{Deltheta2}) which correspond to the green region ($C=1$) and the blue region ($C=-1$), respectively. The process of the topological phase transition induced by the magnetization orientation tuning is clearly presented from (d) to (f) in Fig.~\ref{phase}, corresponding to the origin moving out of the ellipse $O_{1}$ shown in Fig.~\ref{BZ}(b). In addition, according to the bulk-edge correspondence \cite{PhysRevD.13.3398}, there should be $|C|$ chiral edge states on each edge. So it can be inferred that $2|C|$ chiral edge states should be present in the band structure of a strip in Fig.~\ref{phase}(d). Indeed, in the case of $C=1$ we have  two chiral edge states in the band gap. This result shows the reliability of our band structure calculation and the analysis of the topological property.
\section{Numerical results}\begin{figure*}
	\centering
	\includegraphics[scale=0.40]{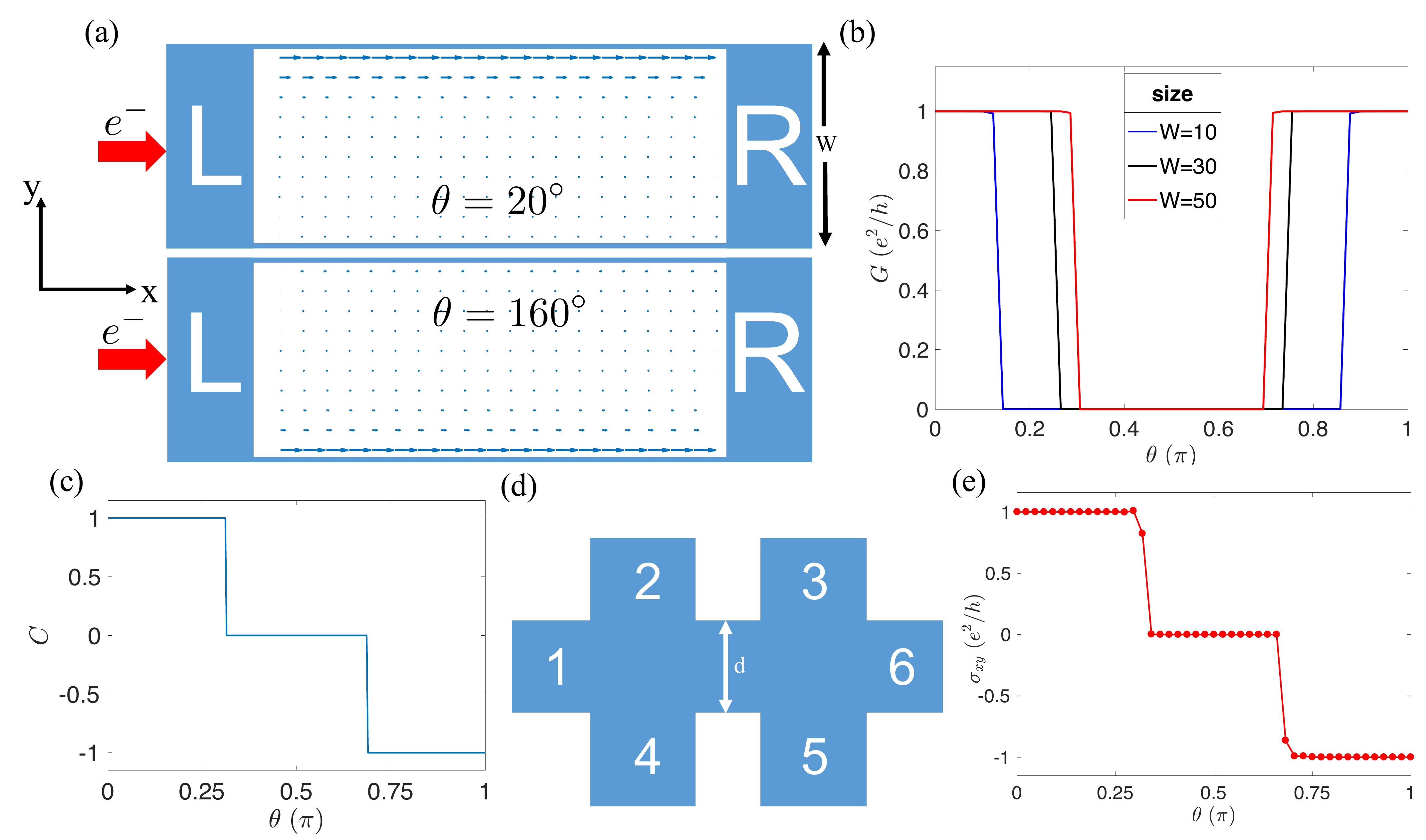}
	\caption{\label{numerical}NEGF calculation results. This set of figures show the numerical results that confirm the analytic ones mentioned in the previous section. In our calculations, parameters are set to be: $t=-0.5\;\text{eV}$, $\Delta=|t|/2$, and $t_{D}=0.6|t|$, the same as the setting of the black solid line in Fig.~\ref{phase}(b). In (a), the cases of $\theta=20^{\circ}$ and $\theta=160^{\circ}$ show opposite chirality, matching the fact that they possess different Chern numbers as shown in Fig.~\ref{phase}(b). In (b), the case with $\text{W}=50$ demonstrates the phase transition points approximately at $\theta=0.31\pi$ and $0.69\pi$, almost the same as the Chern number in (c) extracted from Fig.~\ref{phase}(b). (d) is a six-terminal Hall-bar setup with $d=17$, in which we can calculate the $\sigma_{xy}$. (e) is the relation between $\sigma_{xy}$ and $\theta$. It matches the Chern number in (c).}
\end{figure*} 
In the previous section, we successfully obtained the phase diagrams which show the topological phase transitions with respect to the magnetization orientation $\theta$. The band structure calculations also clearly demonstrate the topological phase transition as predicted by our analytic results. However, all the previous results don't include the consideration of a bias voltage which would be very crucial in experiments. In this section, therefore, we employ the numerical nonequilibrium Green functions (NEGF) to investigate the transport properties of our system in the linear-response regime \cite{Datta_easy}. In the Landauer setup, the sample (central region) is contacted by the left lead and right lead, as shown in Fig.~\ref{numerical}(a). Both the sample and the leads are made up of the system having been in discussion. Therefore, this Landauer setup in our NEGF calculations is a just like the system used to do the band structure calculations in the previous section. But the only difference here is that we rotate our system by $\pi/2$ clockwise in order to match the conventions of conductance or resistance that we are going to discuss later.  
 
In NEGF calculations, one can obtain the lesser Green function $G^{<}(E)$ by using this widely-known formula \cite{Datta_hard}:
\begin{equation}\label{lesser}
	G^{<}(E)=G^{\text{R}}(E)\Sigma^{<}(E)G^{\text{A}}(E),
\end{equation}
where $G^{\text{R}}(E)$ is the retarded Green function, $G^{\text{A}}(E)$ is $[G^{\text{R}}(E)]^{\dagger}$, and $\Sigma^{<}(E)$ is the lesser self-energy. Numerical calculation of the lesser Green function is a very standard technique. One can see the Appendix for relevant information. After knowing the numerical result of $G^{<}(E)$, the physical observable can be obtained through the density matrix
\begin{equation}\label{density_matrix}
	\hat{\rho}=\frac{1}{2\pi i}\int_{E_{F}-\Delta E/2}^{E_{F}+\Delta E/2} G^{<}(E) dE,
\end{equation}
where $E_{F}=-0.01|t|$ is the Fermi energy lying in the bulk gap and $\Delta E=1\times10^{-3}|t|$ is the potential energy drop between two leads with applied bias.
In Fig.~\ref{numerical}(a), we calculate the real-space local charge current flowing from site $\mathbf{m}$ to its nearest neighbor site $\mathbf{m}'$ with the definition \cite{Nikolic_current}
\begin{equation}\label{current}
	J_{\mathbf{m}\mathbf{m}'}=\frac{e}{i\hbar}[c_{\mathbf{m}'}^{\dagger} t_{\mathbf{m}'\mathbf{m}}c_{\mathbf{m}}-\text{H.c.}],
\end{equation} 
where $t_{\mathbf{m}'\mathbf{m}}$ is the $2\times 2$ hopping matrix from $\mathbf{m}$ to $\mathbf{m}'$ that can be found in the real-space Hamiltonian~(\ref{real-space}). One can see that the systems with different magnetization orientations corresponding to different Chern numbers $\pm 1$ can demonstrate opposite chirality. In addition, there exists an unbalance of charge currents between two opposite edges due to the quantum anomalous Hall effect that makes electrons move along the transverse direction. 

In Fig.~\ref{numerical}(b), we calculate the conductance as functions of magnetization orientation $\theta$ at different sizes of the sample $\text{W}$ by this formula \cite{Datta_easy}

\begin{equation}\label{transmission}
	G=\frac{e^{2}}{h} \text{Tr}(\Gamma_{\text{L}} G^{\text{R}} \Gamma_{\text{R}} G^{\text{A}}),
\end{equation} 
where $\Gamma_{\text{L},\text{R}}$ is the level-broadening of the left and right leads, and $\text{Tr}(\Gamma_{\text{L}} G^{\text{R}} \Gamma_{\text{R}} G^{\text{A}})$ is the transmission probability from the left to the right leads. For the definition of level-broadening, one can see the Appendix. In this calculation, the Fermi energy lies in the band gap, so the contribution to the conductance completely comes from the chiral edge states. With the increasing of the sample size (diminishing of the size effect), the conductance $G$ of the case with $\text{W}=50$ as a function of $\theta$ shows the same behavior of topological phase transitions as the Chern number in Fig.~\ref{numerical}(c) extracted from the phase diagram in Fig.~\ref{phase}(b). Therefore, the analytically-obtained phase diagram in the previous section is consistent with the calculation of conductance using NEGF. In addition, the transmission is equal to $|C|$, conforming to the bulk-edge correspondence \cite{PhysRevD.13.3398}.

However, the preceding numerical results are obtained from the two-terminal setup, in which the Hall conductance or resistance can not be calculated. Therefore, we add four more leads contacted to the upper and lower edges of the original two-terminal setup, as shown in Fig.~\ref{numerical}(d). In this Hall-bar geometry, the voltage drops in the longitudinal and transverse directions can be obtained from the Landauer-B\"{u}ttiker formula \cite{Datta_hard}
\begin{equation}\label{Landauer}
	I_{i}=\frac{e^{2}}{h}\sum_{j}(T_{ji}V_{i}-T_{ij}V_{j}),
\end{equation}
where $I_{i}$ is the current flowing from the $i^{\text{th}}$ lead into the sample, $V_{i}$ is the voltage on the $i^{\text{th}}$ lead, and $T_{ji}$ is the transmission from the $i^{\text{th}}$ to the $j^{\text{th}}$ leads. In our NEGF calculation, we can obtain the transmissions $T_{ji}$ between any two arbitrary leads with a voltage applied between the first and sixth leads. Consequently, we can solve Eq.~(\ref{Landauer}) to get the voltage on each lead. The longitudinal and transverse resistances are given by 
\begin{equation}\label{resistance}
	\rho_{xx}=\frac{V_{3}-V_{2}}{I},\;\;\rho_{yx}=\frac{V_{3}-V_{5}}{I},
\end{equation} 
where $I$ is the current injected into the sample from the first lead. Substitute $\rho_{xx}$ and $\rho_{yx}$ into the resistance-to-conductance conversion relations given by
\begin{equation}\label{conversion}
	\sigma_{xx}=\frac{\rho_{xx}}{\rho_{xx}^{2}+\rho_{yx}^{2}},\;\;\sigma_{xy}=\frac{\rho_{yx}}{\rho_{xx}^{2}+\rho_{yx}^{2}},
\end{equation}
we can successfully obtain the $\sigma_{xy}$ whose plateau feature is very crucial in identifying the topologically nontrivial phases. Following the preceding procedure, we employ the NEGF to obtain the Hall conductance $\sigma_{xy}$ as a function of magnetization orientation $\theta$, as shown in Fig.~\ref{numerical}(e). One can easily observe that the numerical result of $\sigma_{xy}$ has a phase transition pattern nicely matching the Chern number in Fig.~\ref{numerical}(c) which comes from the analytic result in the previous section. The proportionality between them conforms to Eq.~(\ref{cond}). Therefore, we can also reach the consistency between the analytic results and transport calculations in the six-terminal Hall bar setup, as what we have done in the standard two-terminal case. For understanding the physical reason of those two non-integer data points in Fig.~\ref{numerical}(e), we also study the cases with the width of the channel $d$ equal to $16$ and $18$, compared to the case of $d=17$ in the figure. Our calculation shows that the non-integer points would move toward or away from the zero plateau when d is $16$ or $18$, respectively. It evidently means that the non-integer conductance originates from the finite-size effect.  
\section{Azimuthal degree of freedom}
\begin{figure}
	\centering
	\includegraphics[scale=0.45]{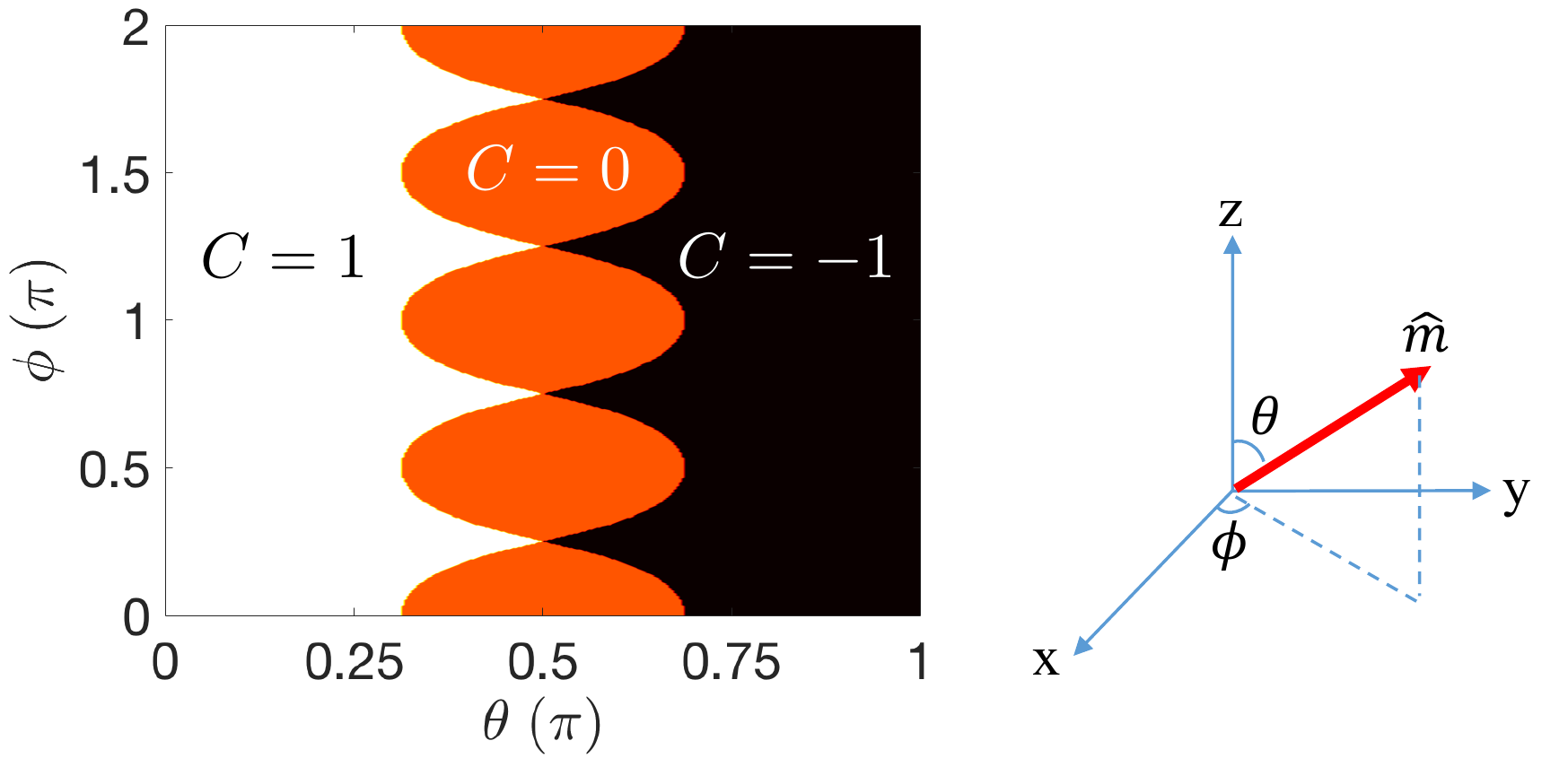}  
	\caption
	{\label{Chern_phase}The setting of the parameters is $t=-0.5$ eV, $\Delta=|t|/2$, and $t_{D}=0.6|t|$. This is the phase diagram in the case that $\theta$ and $\phi$ are both tunable degrees of freedom. The $C=1$, $0$, and $-1$ regions are characterized by white, red, and black, respectively.} 
\end{figure}

So far, we successfully grasp the phase transition behavior of our QAH system in discussion. But remember that the azimuthal angle $\phi$ of the magnetization orientation has been set to zero for convenience in analysis, as mentioned at the Sec. \ref{TPT}. So we relax the degree of freedom $\phi$ and numerically calculate the Chern number given by Eq.~(\ref{Chern}). Fig.~\ref{Chern_phase} is just the phase diagram showing the distribution of the Chern number with respect to $\theta$ and $\phi$. In this phase diagram, topological phase transitions can be induced by tuning $\theta$ at an arbitrary value of $\phi$, not only limited to the case with $\phi=0$. The phase boundary (between $C=1$ and $0$ or $C=0$ and $-1$) shows an oscillating behavior with respect to $\phi$ and its periodicity is $\pi/2$. More specifically, when $\phi$ is an odd multiple of $\pi/4$, the topological phase transition  happens between $C=\pm 1$, without $C=0$. In another aspect, the topological phase transitions can also be induced by tuning $\phi$ at some specific values of $\theta$, if $\theta$ is located within the left and right bounds of the red region ($C=0$). That is to say, if we can somehow make the magnetization precess around the z-axis (\textit{e.g.} microwave \cite{precession,sonhsien-microwave}) with a proper $\theta$ and a constant angular velocity, the topological phase transitions can happen periodically. Such kind of time-dependent dynamics in the topological Floquet system \cite{Gil_Floquet_Nature,Floquet_Chern} is worth further investigation in the future. 
\section{Summary and Discussion}
\begin{figure}
	\centering
	\includegraphics[scale=0.40]{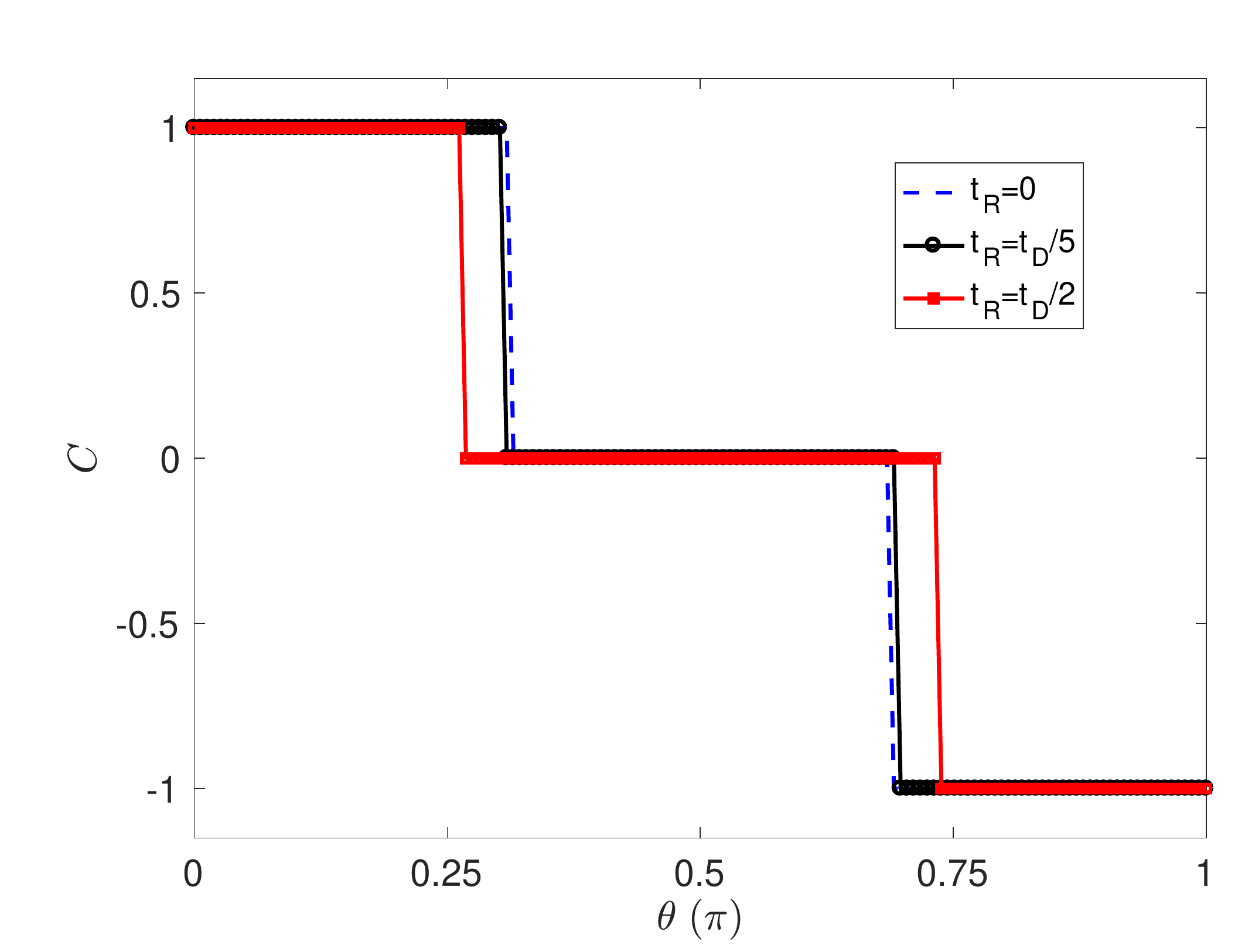}  
	\caption
	{\label{Rashba}As a Rashba term is put into the Hamiltonian, the topological behavior still remains the same, except some shifts in phase transition points. The blue dashed line ($t_{R}=0$) is just the case in Fig.~\ref{numerical}(c).} 
\end{figure}  
We realize a Chern insulator in a 2DES formed by the FM semiconductor/nonmagnetic semiconductor heterostructure. The topological phase transitions in this 2DES system can be induced by tuning the magnetization orientation $\hat{\boldsymbol{m}}$ which is more practical than tuning the exchange coupling strength $\Delta$. The analytic results are highly consistent with the band structure and transport calculations, confirming the validity of this theoretical proposal. Furthermore, 2DESs are very common in today's semiconductor engineering. Therefore, this Chern insulator is feasible to be experimentally realized in a semiconductor heterostructure. However, we think there are still some points worth being discussed. In the following, the satisfaction of the constraint of topological phase transitions, the presence of Rashba SOC, and the possible spontaneous ferromagnetism will be shortly discussed.

By Fig.~\ref{BZ}, we can know that the topological phase transitions are controlled by the interplay between these four parameters, $t_{D}$, $t$, $\theta$, and $\Delta$. In the case of $\theta=\pi/2$, the origin must be out of those two ellipses, resulting a topological trivial phase. Therefore, if the origin can be inside $O_{1}$ when $\theta=0$, allowed by the geometrical constraint of $2|t|>\Delta$, a topological phase transition will definitely happen somewhere during the sweeping of $\hat{\boldsymbol{m}}$ between $\theta=0$ and $\theta=\pi/2$. In experiments, one can adjust the spin-dependent effective mass that determines $t$ by manipulating the carrier density in the semiconductor heterostructure \cite{spin-dependent} for satisfying the constraint of $2|t|>\Delta$. Regarding the parameter $t_{D}$, if it is nonzero under the aforementioned constraint, the topological phase transitions are guaranteed to happen. Thus, with the tunable $t$ and nonzero $t_{D}$, the topological phase transitions can be induced in this 2DES. 

In the transport analysis, the Fermi energy is lying in the band gap such that this system is an insulator. In real cases, however, the Fermi energy may not be lying in the band gap when this system is fabricated. To deal with this problem, one can apply a gate-voltage to tune the Fermi energy into the band gap \cite{QAHexp}. Nonetheless, the Rashba SOC is unavoidably induced when a gate-voltage is applied. In order to understand the effects of Rashba SOC on the topological property, we calculate the Chern number as a function of $\theta$, identical to what has been done in Fig.~\ref{numerical}(c), in the presence of Rashba SOC. The Hamiltonian of Rashba SOC can be easily obtained by replacing $\sigma_{x}(\sigma_{y})$ in $H_{D}$ of Eq.~(\ref{real-space_Dresselhaus}) with $-\sigma_{y}(-\sigma_{x})$. And we use $t_{R}$ (same footing as $t_{D}$) to denote the strength of Rashba SOC. As shown in Fig.~\ref{Rashba}, the cases of nonzero $t_{R}$ show the same behavior of topological phase transitions as the case of zero $t_{R}$, except some shifts in phase transition points. That is to say, the nontrivial topology is still robust in the presence of Rashba SOC.  

According to Ref. \cite{PhysRevB.96.235425}, there is spontaneous ferromagnetism caused by electron-electron interactions in a 2D system with Rashba SOC. Becuase the Rashba SOC and Dresselhaus SOC can be rotated to each other, we think the spontaneous ferromagnetism could also exist in our Chern insulator. That is to say, the required ferromagnetism can be provided by electron-electron interactions so that the FM zinc-blende semiconductor is possible to be replaced with a normal (non-magnetic) zinc-blende semiconductor. Therefore, the fabrication of the semiconductor heterostructure is more realizable. After all, an FM semiconductor is not as common as a non-magnetic semiconductor. This new setup is left as our future work. 
 
\begin{acknowledgments}
We thank M.-C. Chang and S.-Q. Shen for fruitful duscussions. This work is supported by the Ministry of Science and Technology of Taiwan under Grant No. MOST 107-2112-M-002-013-MY3.  
\end{acknowledgments}

\appendix*
\section{Nonequilibrium Green functions (NEGF)}
This is a powerful tool to study the transport properties of electrons in nonequilibrium states, especially for a system applied with a small bias voltage. The essence of NEGF is to calculate the lesser Green function:
\begin{equation}
G^{<}(E)=G^{\text{R}}(E)\Sigma^{<}(E)G^{\text{A}}(E),
\end{equation} 
where
\begin{equation}
G^{R}(E)=[E-H-\sum_{p}\Sigma_{p}(E-eV_{p})]^{-1}.
\end{equation}
$H$ is the Hamiltonian of Eq.~(\ref{real-space}), $\Sigma_{p}$ and $V_{p}$ are the self-energy and voltage of the lead $p$. 
\\The lesser self-energy is
\begin{equation}
\Sigma^{<}(E)=i\sum_{p}\Gamma_{p}(E)f_{p}(E),
\end{equation}
where
\begin{equation}
\Gamma_{p}(E)=i[\Sigma_{p}(E-eV_{p})-\Sigma_{p}^{\dagger}(E-eV_{p})]
\end{equation}
and $f_{p}(E)=f_{0}(E-eV_{p})$ are the level broadening and Fermi-Dirac distribution at zero-temperature of the lead $p$, respectively. Note that $eV_{L}-eV_{R}$ is simply the potential energy drop $\Delta E$ in the upper and lower limits of Eq.~(\ref{density_matrix}). For the calculation of level broadening, one can find the calculation techniques  described in Ref. \cite{Datta_easy}.

\bibliography{QAH}

\end{document}